\documentclass{kapproc} 

\usepackage{procps} 

\usepackage[dvips]{graphicx}

\upperandlowercase

\let\footnotetext\savefootnotetext

\kluwerbib 

\begin{document}

\articletitle[SCUBA Local Universe Galaxy Survey]
{SCUBA Local Universe
Galaxy Survey}

\author{Catherine Vlahakis\altaffilmark{1}, Stephen Eales\altaffilmark{1} and Loretta Dunne\altaffilmark{2}} 
 
\affil{\altaffilmark{1}School of Physics and Astronomy, 
Cardiff University, 5 The Parade, Cardiff, CF24 3YB, UK
\altaffilmark{2}School of Physics and Astronomy, 
University of Nottingham, Nottingham, NG7 2RD, UK}

\begin{abstract}
We discuss the progress of the SCUBA Local Universe Galaxy Survey (SLUGS), the first large, statistical sub-mm survey of the local universe.
	Since our original survey of a sample of 104 \textit{IRAS}-selected galaxies  we have recently completed a sample of 78 Optically-Selected galaxies.
	 Since SCUBA is sensitive to the large proportion of dust too cold to be detected by \textit{IRAS} the addition of this optically-selected sample allows us for the first time to determine the amount of cold dust in galaxies of different Hubble types. We detect 6 ellipticals in the sample and find them to have dust masses in excess of $10^{7}$ $M_{\odot}$. We derive local sub-mm luminosity functions, both directly for the two samples, and by extrapolation from the \textit{IRAS} PSCz, and find excellent agreement.

\end{abstract}

\section{Introduction}
Relatively little is known about the sub-mm properties of ``normal'' galaxies in the local universe -- prior to SCUBA there existed only a handful of sub-mm flux measurements or maps.   SLUGS is the first, large, systematic survey of the local sub-mm universe. It consists of a sample selected from the \textit{IRAS} Bright Galaxy Sample (Dunne et al., 2000), and a sample selected from the CfA optical redshift survey (Vlahakis et al., in prep., Fig. 2). With the optically-selected sample we seek to understand for the first time key questions such as how the amount of cold dust varies with Hubble type.

\section[The $850\mu$m Luminosity Function]
{The $850\mu$m Luminosity Function}
Using the \textit{IRAS}-selected sample Dunne et al. produced the first direct estimate of the sub-mm luminosity function (LF). However, since the \textit{IRAS} sample is biased toward galaxies with larger amounts of warmer dust its LF may also be subject to bias. Conversely, the optically-selected sample should, by definition, be free from temperature selection effects. 

Here,  we derive a direct $850\mu$m LF for the optically-selected sample. However, in order to better constrain the LF at the lower luminosity end we need more data points, spanning a greater range of luminosities. We do this by determining an $850\mu$m LF using $\sim10000$ galaxies from the \textit{IRAS} PSCz survey. We predict their $850\mu$m luminosities by extrapolating their \textit{IRAS} fluxes using the colour-colour relation from SLUGS (Serjeant and Harrison, 2003). In Fig. 1 we compare the extrapolated (PSCz) and direct LFs and find excellent agreement between all 3 LFs.

\begin{figure}[ht]
\vskip.2in
\sidebyside
{\includegraphics[angle=270,width=2.5in]{vlahakis_fig01.ps}}
{\includegraphics[angle=270,width=1.9in]{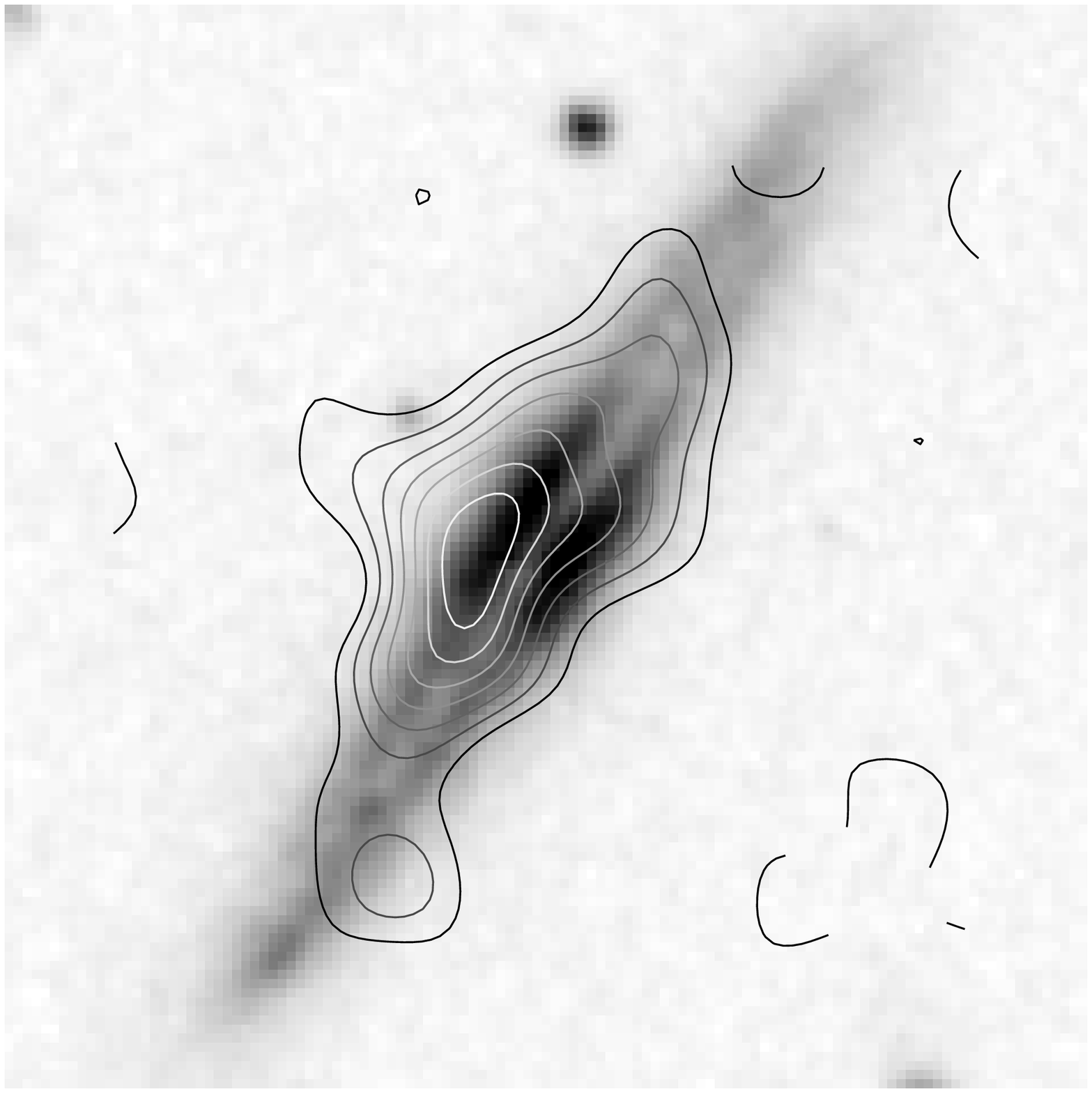}}
\sidebyside
{\caption{$850\mu$m LF projected from the PSCz, compared with directly measured LF from optically- and \textit{IRAS}-selected samples (circles, stars, triangles respectively).}}
{\caption{Example from the optically-selected SLUGS: NGC 3987. 850 $\mu$m S/N map ($1\sigma$ contours) overlaid onto \textit{Digitised Sky Survey} optical image.}}
\end{figure}

\section[Ellipticals in the Optically-Selected SLUGS]
{Ellipticals in the Optically-Selected SLUGS}

It was once thought that ellipticals were entirely devoid of dust and gas, but optical absorption studies now show that dust is usually present.  Dust masses for the $\sim15\%$ of ellipticals detected by \textit{IRAS} have been found to be as much as a factor of 10--100 higher when estimated from their FIR emission compared to estimates from optical absorption, suggesting a diffuse cold dust component (Goudfrooij and de Jong, 1995 and refs. therein, Bregman et al., 1998). 
At  $850\mu$m we detect 6 ellipticals, from a total of 11 ellipticals in the optically-selected sample, and find them to have dust masses in excess of $10^{7}$ $M_{\odot}$. We will investigate this further with SCUBA observations of a larger sample of ellipticals.

\end{document}